# Machine Learning and Artificial Intelligence in Next-Generation Wireless Network

Wafeeq Iqbal, Wei Wang, Ting Zhu

University of Maryland, Baltimore County

**Abstract-** Due to the advancement in technologies, the next-generation wireless network will be very diverse, complicated, and according to the changed demands of the consumers. The current network operator's methodologies and approaches are traditional and cannot help the next generation networks to utilize their resources most appropriately. The limited capability of the traditional tools will not allow the network providers to fulfill the demands of the network's subscribers in the future. Therefore, this paper will focus upon "machine learning", automation, "artificial intelligence", and "big data analytics" for improving the capacity and effectiveness of next-generation wireless networks. The paper will discuss the role of these new technologies in improving the service and performance of the network providers in the future. The paper will find out that machine learning, big data analytics, and artificial intelligence will help in making the "next-generation wireless network" self-adaptive, self-aware, prescriptive, and proactive. At the end of the paper, it will be provided that future wireless network operators cannot work without shifting their operational framework to AI and machine learning technologies.

*Ker terms- Artificial* Intelligence, Big data, wireless networks, machine learning.

## 1. Background

Machine learning and big data analytics are playing the most important role in the Next-generation wireless networks (NGWN) [12-46]. NGWN is a dynamic and complex network and it included the fifth generation and beyond networks. The next generation is service driven and there is a huge need that a single infrastructure provided multiple services to increase the efficiency and effectiveness of the system. The diverse services which are provided by NGWN include low latency, massive type and ultra-reliable communication, and increased mobile broadband. NGWN should be capable of ensuring access to multiple services such as long-term evolution, Wi-Fi, and fifth generation. Despite providing multiple services, NGWN should be capable of dealing with the different nature of networks that belonged to various kinds of base stations such as micro, macro, pico BSs, Femto user's applications, and devices [1]. For any network operator to operate with efficiency and to satisfy the users by providing diverse and flexible services is an enormous challenge. One of the major issues for the network operator in terms of providing their services to the diverse number of users included the extending and more efficient network services with the help of the limited resources and capacity pool. The network operator needs to fulfill the growing needs of consumers for the capacity and efficiency of the service in terms of the network. Machine learning and artificial intelligence will help in the advanced planning and management of these diversified networks because manual dealing with these networks will leave the system in more complicated situations. Moreover, manual operating of the networks or the interaction of human beings with the machine can increase the chances of errors and result in more costly and time-consuming processes [2].

Artificial intelligence has the potential or the ability of computers or robots to command the system for the work and perform all the works that can be done with human intelligence. Artificial intelligence allows the network operators to plan, perform and manage diverse services to fulfill the increasing demands of the consumers over time. Machine learning is a specific form of Artificial intelligence that supports the network operators in planning by providing the accurate prediction about the results of the programs without their actual implementations [3]. Machine learning and artificial intelligence will help the system to automate the process and enhance the level of efficiency within the minimum operating cost. The next-generation wireless networks can be 5G or beyond and these networks cannot work properly without implementations of automation.

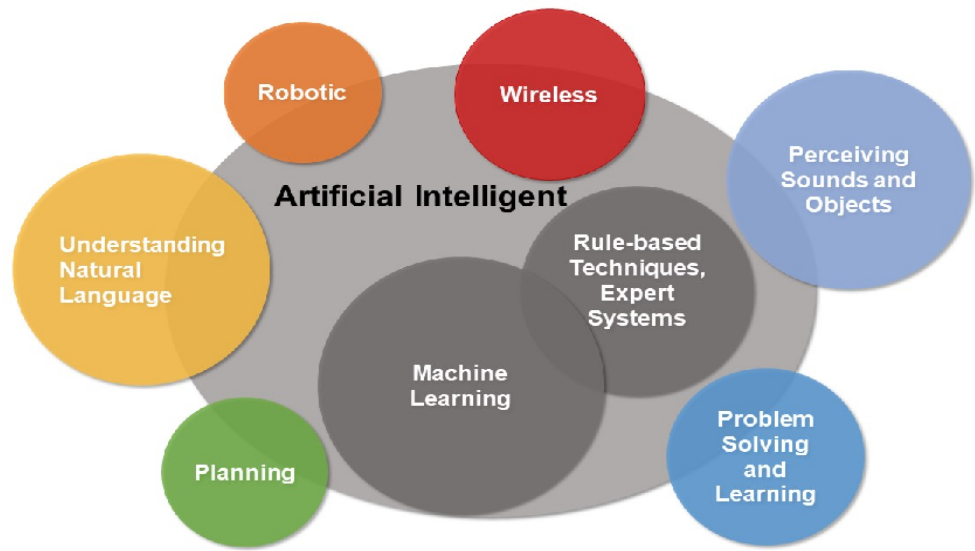

Figure.1. The relationship between machine learning and artificial intelligence that can improve the outcome of the next-generation wireless network operations

Network operators use various approaches to optimize their services and the most effective approach which they are using even in the present day is the optimization of "single key performance indicators (KPIs)" which can only work with less amount of data. KPIs cannot work with large sources of data because it is capable of optimizing one element in the network operators in a specific period [4].

At the current time, the network operators are utilizing the old data record and deploying traditional tools on the various parts or locations of data networks that depend upon accumulated KPIs. The use of outdated data is negatively impacting the capacity of the network operators. Generally the current and the next generation network operators have to deal patterns and relations and can help in providing a more accurate understanding of the data that is collected from multiple sources. In this way, big data analysis and artificial intelligence will help in providing with a large number of users and the networks. Artificial intelligence is capable of operating unknown detailed knowledge and understanding of the values and ensure effective measures to improve the effectiveness of the overall wireless networks [5].

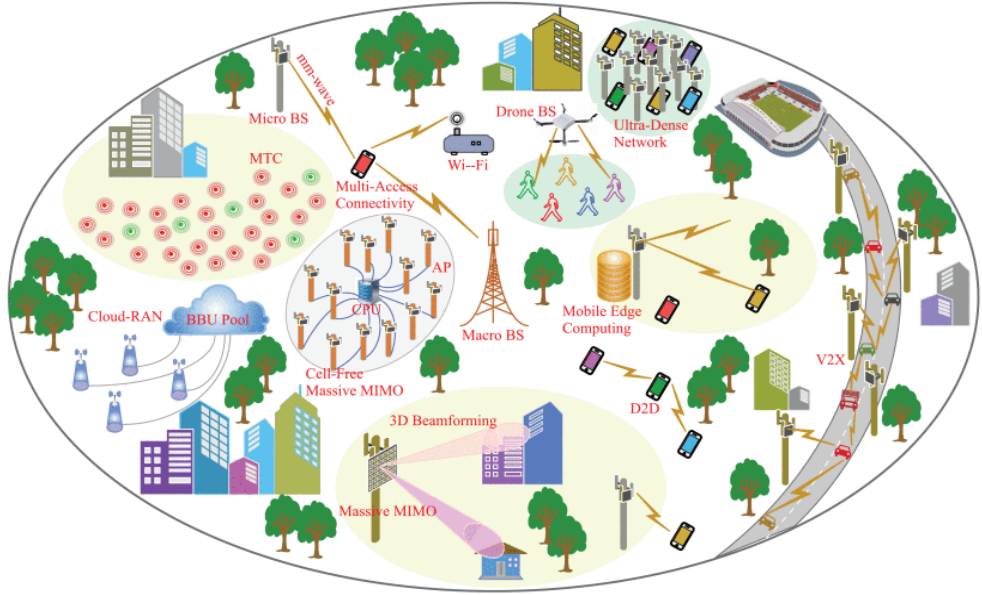

Figure. 2. The graphical representation of a few of the technologies and elements that will combine to form the next-generation wireless communication networks is shown in the figure.

Big data and artificial intelligence provide value to the networks with the help of including the large number of data sources that utilized the quality of experience and the customer's centered approaches for enhancing the quality and performance of the next-generation wireless network operators. "Experimental network intelligence (ENI)" is a specific group that helps in developing effective network architecture, and it is based upon the technology of artificial intelligence. ENI is created by "The European telecommunications standards institute (ETSI)" and proving very beneficial for third-generation wireless networks. There is the possibility that the performance of the future generations of the network can be improved with the help of utilizing the maximum elements of the ENI model [6]. ENI model will not only in the enhancing the configuration of the system but also optimize the process of monitoring the services.

Artificial intelligence will not only impact the next generation wireless system operation in terms of the operational framework but it will also affect the operational expenses of the networks. If the future network operator needs to fulfill the growing demands of consumers there is a huge need that they should be self-aware, smart, autonomous, and self-adaptive and manage the system cost-effectively [7]. For the future more diverse and advanced networks, the conventional management and maintenance approaches cannot work effectively. Using big data analytics can help future networks operators to provide the best predictive and proactive elements and overcome all the challenges that the current network systems are facing because of the use of some manual and conventional approaches to deal with the data. With the help of big data analytics, the system can go beyond only maintaining it because it will help in involving a large volume of data, the maximum number of data sources and allows best predictions that will help in taking effective decisions that can save the loss of time and money of the network operators. Machine learning and the tools of artificial intelligence are capable of uncovering the wireless network operators' properties including, anomalies, correlation, and unknown characteristics that are not possible to know with the help of manual approaches and inspections. Getting knowledge and understanding of the unknown properties will help in taking the best decisions to improve the operations and deployment of the next-generation wireless network systems.

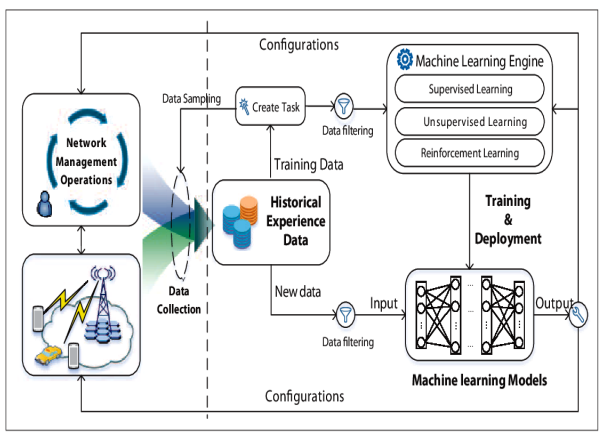

Figure. 3. An effective example of an auto-learning network to get the automation of the network processes with the help of machine intelligence for the wireless network communication is provided in the image.

*a. Big data analytics and NGWN*

The complex patterns of the network traffic and the increasing complications of the wireless system are attracting more towards big data analytics. In the initial days of big data analytics, the network systems were very conscious to shift their operational frameworks but later one the network operators have realized the importance of artificial intelligence for optimizing the efficiency and performance of the network in the near future.

As the result of growing optimization because of big data, the network operators start in-depth understanding and knowledge of this new and most advanced approach towards increasing the performance of the networks. As the result of the search, it is found that there are three most significant factors which attract the network operators toward adopting big data analytics or the approach based upon artificial intelligence. Following is the discussion of the drivers for artificial intelligence in the next-generation wireless networks system.

*1. Services and expenses*

The users of the wireless networks are demanding high-speed data without increasing the cost of the network. They wanted to get the demanded next-generation wireless network within minimum budgets. In these situations, network operators need a framework that can help them to increase the capacity of their system at less cost. Due to increased competition in the market, the network operators need to shift their focus from the network center to the customers' center so that the satisfaction of the consumers can be enhanced. The network operator needed to improve their services, get profit margin and reduce the cost of the operations as much as possible; therefore it gets attracted towards the technology of artificial intelligence and machine learning.

### 2. Usage of network

With the advancement in technology, users are using the network on various kinds of devices or tools. There are different patterns of network usage due to differences in the devices and usage of the users. The traffic of the network systems is increasing as a result of the higher use of technology and data. The network operators need to understand the heterogeneous patterns of the user's data use and deal with the large numbers of traffic in a way that the services of the network and usage of the users cannot be interpreted. The use of big data analytics will allow the network operating systems to effectively use the maximum resources and ensure high-quality services to the traffic.

### 3. Use of technology

The next-generation wireless network included various kinds of users' cases such as edge computing, network slicing, and vitalization. The use of big data analytics can allow the network operators to divide the cases and the network slicing into multiple layers and provide the best services to the consumer. This is not possible without artificial intelligence because of technological advancements and increased network traffic [6].

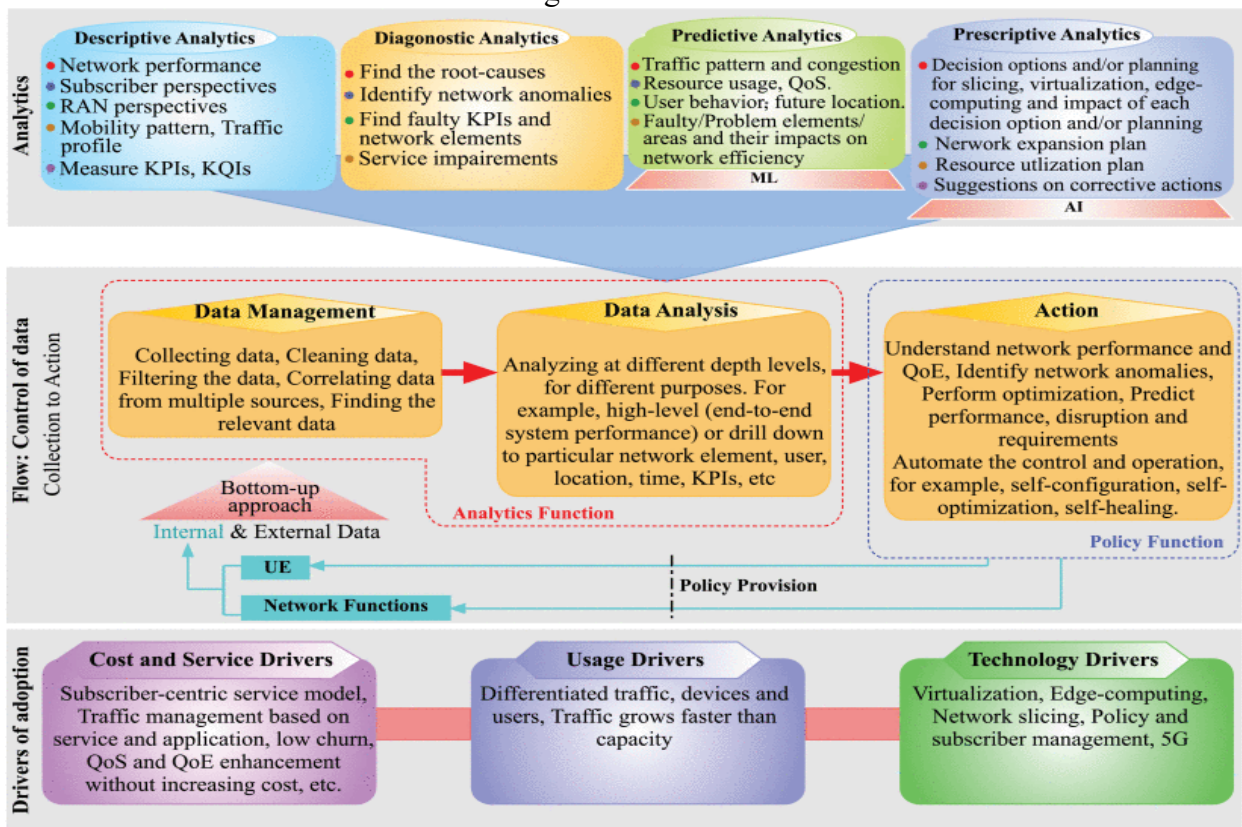

Figure. 4. All the important factors or derivers that help in understanding that big data analytics should be used in the next-generation wireless networks.

b. *Network operators and the data analytics*

Different kinds of data analytics can be used by the network operators to ensure the services to the subscribers. Many diagnostic tools help in detecting the causes of the network operations and errors so that the best decision can be taken for optimizing the performance of the data. Predictive, descriptive, and prescriptive are the most important kinds of data analytics. The network operator systems in the current time are using only the descriptive kind of data analytics and it does not process a large amount of information. Descriptive analytics included the use of visualization tools to inspect the performance and others things related to wireless networks operations. Network operators can make the best predictors with the help of predictive data analytics. It allows the network operators to forecast and what will happen to the traffic, networks, and other elements of the network systems. It included the techniques such as data mining and machine learning so that the best forecasting can be possible for the network operators. Prescriptive analytics is the most advanced data analytics form as it goes beyond prediction because it provided the most significant options for decisions related to vitalization, slicing, and edge computing [8].

c. *External and internal data*

There are two main classes of data which is accessed by the network operating systems and these are external and internal. All the users and network-related data will include in the internal kind of data. The data that is gathered from the third parties will name as external data. These two categories are further divided into unstructured and structured kinds of data.

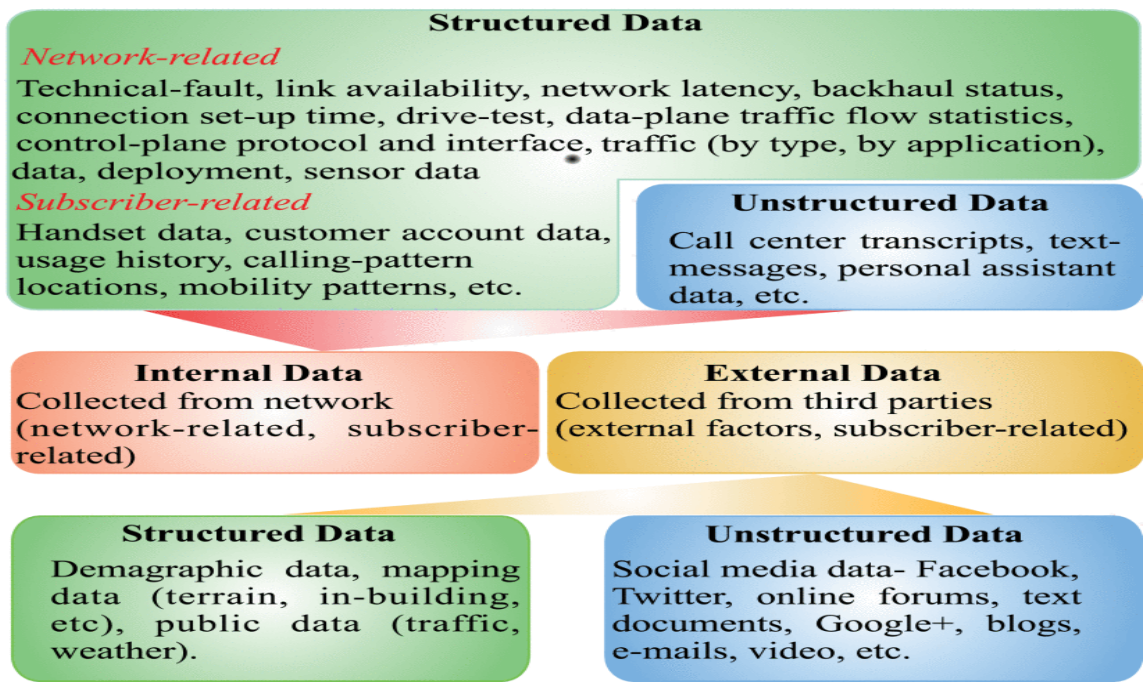

Figure.5. These are a few data sets that are required to the artificial intelligence, machine learning, and big data analytics for the processing of its functions.

## 2. Effectiveness of the Artificial In Intelligence

In the "next generation wireless networks," the network system operators will need to deal with multiple kinds of data, for example, unpredicted, unstructured, and many others. The use of traditional methods or approaches cannot prove helpful for the network operators with complex and heterogeneous data. Artificial intelligence is playing a major role in unveiling the hidden patterns of data or performing big data analytics. Artificial intelligence will help the operating and analyzing the past data to make decisions in real-time [9]. In short, the use of computational intelligence will help the network operators to get meaningful information from the unstructured or the raw data that is collected from multiple resources. There are many kinds of approaches and methodologies that can be applied with the help of artificial intelligence to perform big data analytics. A few of the most important methodologies of big data analytics are stochastic algorithms, rough sets, evolutionary computing, neural algorithms, learning theory, fuzzy logic, swarm intelligence, probabilistic methods, physical algorithms, and immune algorithms [9].

## 3. Big Data Analytics Approaches

Bottom-up and Top-down are two main approaches that are used by the network operating system to perform big data analytics. The bottom-up approach helps in creating novel business opportunities and deals with the RAN and the consumers' perspectives. However, the Top-down approach deals with the resolution of the problems. It will focus on what is problem, what kind of data will be required to address it, and how it could be solved effectively to improve the overall performance and the service of network operators [10].

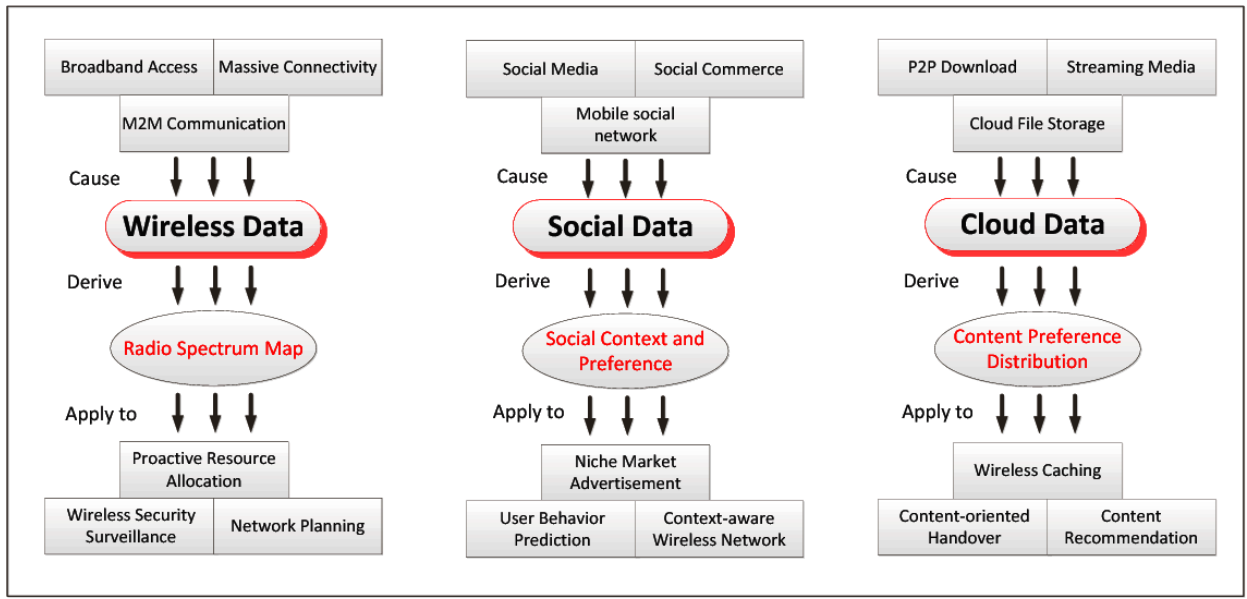

Figure.6. The sources of the wireless big data, its applications, and the hidden information in the networks are shown in the figure [12].

Artificial intelligence and Machine learning can process complex data The two most important tools for dealing with a large amount of complex data, for predicting the information, and forecasting for the best solutions to the problems are artificial intelligence and machine learning. In "machine learning", it is possible sometimes that the command and access to enormous data are provided to the machine and it will work like artificial intelligence. Contrary to this, artificial intelligence included the devices that take decisions themselves by performing intelligent tasks. For the next-generation wireless network, applied artificial intelligence can be proved more effective as compared to its generalized form because applied can help in dealing with and controlling the mobile networks. The combination of artificial intelligence is best with machine learning for the future generation of wireless networks because machine learning helps network operators to predict the data and future patterns while artificial intelligence help in suggesting and prescribing the solutions to the problems[5].

AI and machine learning are very helpful for the changing technologies of networks because it does not only help in providing meaning to the raw data but also allow in operating large amounts of complex operations. It can deal with multiple sources and prove effective in the new areas where there is no previous data and artificial intelligence need to prescribe the solutions without analyzing the previous data [2].

No doubt, human beings are very intelligent and their level of expertise can enhance the performance of the network operators but the human beings have limited capacity and the chances of errors are higher therefore the future of wireless networks is associated only with the approaches and tools that deal with the artificial intelligence [8].

It is proved from the study that there is a large number of challenges and issues faced by the future generation wireless networks that cannot be addressed with the help of traditional approaches. If the network operators try to deal with the next-generation wireless networks with the help of old technology the result will be low customer satisfaction and a lack of proper wireless network management. But it is highly recommended that all the future challenges involving the issues of multiple sources, huge traffic, user access, and many others can be easily solved with the help of artificial intelligence [8].

### 4. Increased Capacity of the NGWN

Big data analytics and AI help increase the capacity of network operators with the help of providing predicting the data most effectively. The traditional approaches cannot prove helpful in these situations for example the mobile network operator may require some solutions to resolve the issues and because of it, the big data analytic will be able to get all the information at the same place so that the error can be detected and resolved with the minimum time duration. The users' and the network's perspective data can be analyzed and predicted with the help of using big data analytics.

The data of the subscriber or the user's perspective will include the affordability of the users, data of the user's device, policies, quality of the services, and behavioral data. The analysis of the user's data will help in taking the user's center decision so that maximum outcome can be obtained in terms of increasing the customers' satisfaction.

The network perspective data will include all the processes which will include planning, managing, and monitoring the processes, determining the pricing, dealing with the complex situation, and resolving the issues of the network operating system so that the efficiency of the performance can be increased.

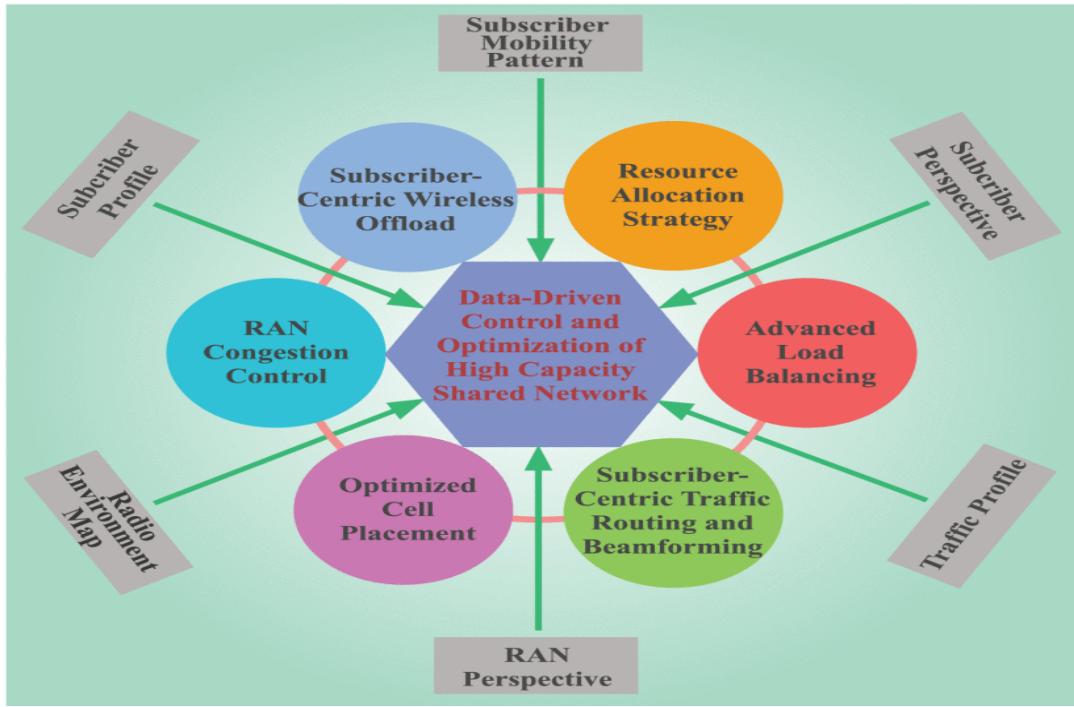

Figure.7. The graphical representation of different data analytical applications and how these can help in enhancing the effectiveness in the next-generation wireless networks.

There are different ways in which the communication process can be improved for the next-generation wireless networks. Following are the detailed description of the grew most effective that can be employed by the next-generation networks.

1. With the help of accurate resources allocation process, it is possible to improve the future generation communication process. It has been reported that using big data analytics for the proper allocation of resources can prove very advantageous for future generations. For example with the help of using data analytics, the network operators can plan and manage their resources according to the demand of the users and it will be beneficial for the future wireless system to ensure effective services to the users.
2. It is very important to provide users with the quality of experience and it is only possible with the help of big data because the data will allow in selecting the best solutions according to the user's preferences and the kind of application that are used by them.
3. With the help of big data and the intelligence of machines, the network operators will be able to deal with the large surge of traffic. By using the data-driven technologies and approaches the network operators can effectively decide which small device or cell phone will be connected with the networks according to the user's affordability and the profile.
4. Placing small cells can help the network operators to increase the capacity and efficiency of the network but it is very hard for the

network operators to find out the best or most favorable location for the placement of small cells. However, machine learning or big data analytics can prove most effective in this regard because it can best predict even at new locations.

5. The growing demands of the users for the networks and the limited resources of the network operators can lead to the "Radio Access Network Congestion Control" that negatively impacted the quality of the services provided to the users. However, the network operators can't use the traditional approach and increase the number of resources because the system will become complex and the result will be even low quality. Thus it is assumed that artificial intelligence and big data analytics can best help the network operators in such situations.

6. Each device or cell phone has different traffic and the profile that resulted in the creation of distinctive data and patterns for each cell. When some subscribers get disconnected with particular cells and get connected with the others it increases the load upon the other and as a result, the traffic and patterns get disturbed. Predictive analysis with the help of data analytic will prove helpful in finding out the location of overload so that the network operators manage the next-generation wireless network effectively in near future.

7. In the "next generation wireless networks", the data rates and coverage of the network services can best be provided with the help of a component of integrated technology named beamforming. Artificial intelligence, big data analytics, and machine learning are the most important techniques which can help in improving the data and coverage rate with the help of detecting the best position and location of antennas. Predicting the load and prescribing the best solution to address the issue.

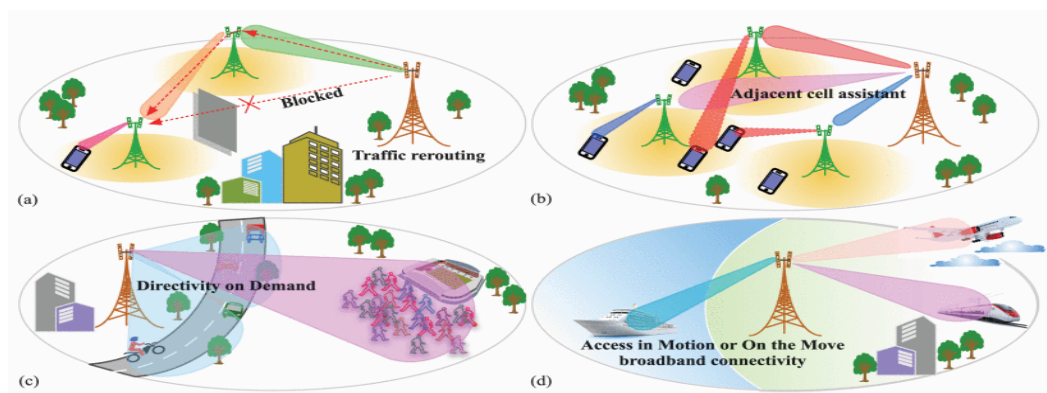

Figure.8. Graphical representation of how different latest technologies can help in beamforming and adjusting in the next-generation wireless networks.

### 5. Limitations

No doubt using big data, artificial intelligence, and machine learning will be very advantageous for wireless networks but many challenges can be faced by the network operators as the result of implementing this most advanced technology into their systems. In the first place, it is very challenging to give machines access to a large number of data resources, and thus it is a challenge itself. Furthermore, there is a need to have a highly-skilled, trained, and smart workforce that is capable of working with the machine. The network operators need to hire new employees and give various kinds of training to the previous employees which can increase the cost of the network operating systems. Next, due to the automation and machine access to the data, there will be a loss of direct control over the data. Despite these challenges it is fact that future generation wireless networks cannot work without automation, artificial intelligence, and big data analytics because of the complexity of the future network operator systems.

## 6. Conclusion

From the above discussion and evidence, it is clear that artificial intelligence, machine learning, and data analytics can prove the best technologies to deal with the next-generation wireless network operators. These technologies allow the network operators to effectively control, manage and improve their performance so that the consumers' satisfaction can be improved by increasing the capacity of the network without increasing its cost.

## REFERENCES

[1]. Boccardi, F., Heath, R.W., Lozano, A., Marzetta, T.L. and Popovski, P., 2014. Five disruptive technology directions for 5G. *IEEE communications magazine*, *52*(2), pp.74-80.

[2]. Acemoglu, D. and Restrepo, P., 2019. *8. Artificial Intelligence, Automation, and Work* (pp. 197-236). University of Chicago Press.

[3]. Das, S., Dey, A., Pal, A. and Roy, N., 2015. Applications of artificial intelligence in machine learning: review and prospect. *International Journal of Computer Applications*, *115*(9).

[4]. Paolini, M., 2017. Mastering analytics: how to benefit from big data and network complexity. *online]. http://content. rcrwireless. com/20170620 Mastering Analytics Report*.

[5]. Bi, S., Zhang, R., Ding, Z. and Cui, S., 2015. Wireless communications in the era of big data. *IEEE communications magazine*, *53*(10), pp.190-199

[6]. Kibria, M.G., Nguyen, K., Villardi, G.P., Zhao, O., Ishizu, K. and Kojima, F., 2018. Big data analytics, machine learning, and artificial intelligence in next-generation wireless networks. *IEEE access*, *6*, pp.32328-32338.

[7]. Han, S., Chih-Lin, I., Li, G., Wang, S. and Sun, Q., 2017. Big data enabled mobile network design for 5G and beyond. *IEEE Communications Magazine*, *55*(9), pp.150-157.

[8]. Engelbrecht, A.P., 2007. *Computational intelligence: an introduction*. John Wiley & Sons.

[9]. Pedrycz, W., Sillitti, A. and Succi, G., 2016. Computational intelligence: an introduction. In *Computational Intelligence and Quantitative Software Engineering* (pp. 13-31). Springer, Cham.

[10]. Acker, O., Blockus, A. and Pötscher, F., 2013. Benefiting from big data: A new approach for the telecom industry. *Strategy&, Analysis Report*.

[11] Chen, M., Challita, U., Saad, W., Yin, C. and Debbah, M., 2017. Machine learning for wireless networks with artificial intelligence: A tutorial on neural networks. *arXiv preprint arXiv:1710.02913*, *9*.

[12] Liu, Y., Bi, S., Shi, Z. and Hanzo, L., 2019. When machine learning meets big data: A wireless communication perspective. *IEEE Vehicular Technology Magazine*, *15*(1), pp.63-72.

[13] Degroote, L., De Bourdeaudhuij, I., Verloigne, M., Poppe, L., & Crombez, G. (2018). The Accuracy of Smart Devices for Measuring Physical Activity in Daily Life: Validation Study. *JMIR mHealth and uHealth*, *6*(12), e10972. https://doi.org/10.2196/10972

[14] B. Reeder and A David, "Health at hand: A systematic review of smart watch uses for health and wellness," in Journal of Biomedical Informatics,, Volume 63, 2016, pp. 269-276. https://www.sciencedirect.com/science/article/pii/S1532046416301137

[15] Chi, Zicheng, et al. "Leveraging ambient lte traffic for ubiquitous passive communication." Proceedings of the Annual conference of the ACM Special Interest Group on Data Communication on the applications, technologies, architectures, and protocols for computer communication. 2020.

[16] Liu, Xin, et al. "Vmscatter: A versatile {MIMO} backscatter." 17th {USENIX} Symposium on Networked Systems Design and Implementation ({NSDI} 20). 2020.

[17] Liu, Xin, et al. "Verification and Redesign of {OFDM} Backscatter." 18th {USENIX} Symposium on Networked Systems Design and Implementation ({NSDI} 21). 2021.

[18] Chi, Zicheng, et al. "Harmony: Exploiting coarse-grained received signal strength from IoT devices for human activity recognition." *2016 IEEE 24th International Conference on Network Protocols (ICNP)*. IEEE, 2016.

[19] Chi Z, Yao Y, Xie T, et al. EAR: Exploiting uncontrollable ambient RF signals in heterogeneous networks for gesture recognition[C]//Proceedings of the 16th ACM conference on embedded networked sensor systems. 2018: 237-249.

[20] Yao, Y., Li, Y., Liu, X., Chi, Z., Wang, W., Xie, T., & Zhu, T. (2018, April). Aegis: An interference-negligible RF sensing shield. In *IEEE INFOCOM 2018-IEEE conference on computer communications* (pp. 1718-1726). IEEE.

[21] Chi, Zicheng, et al. "Countering cross-technology jamming attack." Proceedings of the 13th ACM Conference on Security and Privacy in Wireless and Mobile Networks. 2020.

[22] Chi, Zicheng, et al. "Parallel inclusive communication for connecting heterogeneous IoT devices at the edge." *Proceedings of the 17th conference on embedded networked sensor systems*. 2019.

[23] Chi, Zicheng, et al. "PMC: Parallel multi-protocol communication to heterogeneous IoT radios within a single WiFi channel." *2017 IEEE 25th International Conference on Network Protocols (ICNP)*. IEEE, 2017.

[24] Gu S, Yi P, Zhu T, et al. Detecting adversarial examples in deep neural networks using normalizing filters[J]. UMBC Student Collection, 2019.

[25] Jiang, Dingde, et al. "A traffic anomaly detection approach in communication networks for applications of multimedia medical devices." *Multimedia Tools and Applications* 75.22 (2016): 14281-14305.

[26] Li Y, Zhu T. Gait-based wi-fi signatures for privacy-preserving[C]//Proceedings of the 11th ACM on asia conference on computer and communications security. 2016: 571-582.

[27] Li Y, Zhu T. Using Wi-Fi signals to characterize human gait for identification and activity monitoring[C]//2016 IEEE First International Conference on Connected Health: Applications, Systems and Engineering Technologies (CHASE). IEEE, 2016: 238-247.

[28] Liu Y, Xia Z, Yi P, et al. GENPass: A general deep learning model for password guessing with PCFG rules and adversarial generation[C]//2018 IEEE International Conference on Communications (ICC). IEEE, 2018: 1-6.

[29] Tao Y, Xiao S, Hao B, et al. WiRE: Security Bootstrapping for Wireless Device-to-Device Communication[C]//2020 IEEE Wireless Communications and Networking Conference (WCNC). IEEE, 2020: 1-7.

[30] Wang W, Liu X, Yao Y, et al. Crf: Coexistent routing and flooding using wifi packets in heterogeneous iot networks[C]//IEEE INFOCOM 2019-IEEE conference on computer communications. IEEE, 2019: 19-27.

[31] Wang W, Xie T, Liu X, et al. Ect: Exploiting cross-technology concurrent transmission for reducing packet delivery delay in iot networks[C]//IEEE INFOCOM 2018-IEEE Conference on Computer Communications. IEEE, 2018: 369-377.

[32] Xiao, Sheng, et al. "Reliability analysis for cryptographic key management." *2014 IEEE International Conference on Communications (ICC)*. IEEE, 2014.

[33] Yang Q, Fu H, Zhu T. An optimization method for parameters of SVM in network intrusion detection system[C]//2016 International Conference on Distributed Computing in Sensor Systems (DCOSS). IEEE, 2016: 136-142.

[34] Yi P, Guan Y, Zou F, et al. Web phishing detection using a deep learning framework[J]. Wireless Communications and Mobile Computing, 2018, 2018.

[35] Yi P, Yu M, Zhou Z, et al. A three-dimensional wireless indoor localization system[J]. Journal of Electrical and Computer Engineering, 2014, 2014.

[36] Yi P, Zhu T, Liu N, et al. Cross-layer detection for black hole attack in wireless network[J]. Journal of Computational Information Systems, 2012, 8(10): 4101-4109.

[37] Yi P, Zhu T, Ma J, et al. An Intrusion Prevention Mechanism in Mobile Ad Hoc Networks[J]. Ad Hoc Sens. Wirel. Networks, 2013, 17(3-4): 269-292.

[38] Yi P, Zhu T, Zhang Q, et al. A denial of service attack in advanced metering infrastructure network[C]//2014 IEEE International Conference on Communications (ICC). IEEE, 2014: 1029-1034.

[39] Yi P, Zhu T, Zhang Q, et al. Puppet attack: A denial of service attack in advanced metering infrastructure network[J]. Journal of Network and Computer Applications, 2016, 59: 325-332.

[40] Zhang S, Zhang Q, Xiao S, et al. Cooperative data reduction in wireless sensor network[J]. ACM Transactions on Embedded Computing Systems (TECS), 2015, 14(4): 1-26.

[41] Zhu T, Xiao S, Ping Y, et al. A secure energy routing mechanism for sharing renewable energy in smart microgrid[C]//2011 IEEE International Conference on Smart Grid Communications (SmartGridComm). IEEE, 2011: 143-148.

[42] Zhu T, Xiao S, Zhang Q, et al. Emergent technologies in big data sensing: a survey[J]. International Journal of Distributed Sensor Networks, 2015, 11(10): 902982.

[43] Zhu T, Yu M. A dynamic secure QoS routing protocol for wireless ad hoc networks[C]//2006 IEEE Sarnoff Symposium. IEEE, 2006: 1-4.

[44] Sui Y, Yi P, Liu X, et al. Optimization for charge station placement in electric vehicles energy network[C]//Proceedings of the Workshop on Smart Internet of Things. 2017: 1-6.

[45] Xu J, Yi P, Xie T, et al. Charge station placement in electric vehicle energy distribution network[C]//2017 IEEE International Conference on Communications (ICC). IEEE, 2017: 1-6.

[46] Huang C, Yi P, Zou F, et al. CCID: Cross-Correlation identity distinction method for detecting shrew DDoS[J]. Wireless Communications and Mobile Computing, 2019.